# Analysis of Habitability and Stellar Habitable Zones from Observed Exoplanets

Jonathan H. Jiang[1], Philip E. Rosen[2], Christina X. Liu[3], Qianzhuang Wen[4], Yanbei Chen[5]

[1] Jet Propulsion Laboratory, California Institute of Technology, Pasadena, California, USA
[2] Independent Researcher, Vancouver, Washington, USA
[3] Lakeside School, Seattle, Washington, USA
[4] Pacific Academy, Irvine, California, USA
[5] Burke Institute for Theoretical Physics, California Institute of Technology, Pasadena, CA 91125

Correspondence: Jonathan.H.Jiang@jpl.nasa.gov

## Abstract

The investigation of exoplanetary habitability is integral to advancing our knowledge of extraterrestrial life potential and detailing the environmental conditions of distant worlds. In this analysis, we explore the properties of exoplanets situated with respect to circumstellar habitable zones by implementing a sophisticated filtering methodology on data from the NASA Exoplanet Archive. This research encompasses a thorough examination of 5,595 confirmed exoplanets listed in the Archive as of March 10$^{th}$, 2024, systematically evaluated according to their calculated surface temperatures and stellar classifications of their host stars, taking into account the biases implicit in the methodologies used for their discovery. Our findings elucidate distinctive patterns in exoplanetary attributes, which are significantly shaped by the spectral classifications and mass of the host stars. The insights garnered from our study not only enhance the existing models for managing burgeoning exoplanetary datasets, but also lay foundational groundwork for future explorations into the dynamic relationships between exoplanets and their stellar environments.

## 1. Introduction

The pursuit of understanding the cosmos beyond our own Solar System has revolutionized the field of exoplanetary science, propelled by significant advancements in observational technologies and data analysis methods. The NASA Exoplanet Archive which, as of mid-2024, catalogues nearly 5,700 confirmed exoplanets, serves as a cornerstone for such explorations, providing an expansive database that reflects the diverse characteristics of these distant worlds (NASA Exoplanet Archive, 2024). The remarkable surge in exoplanet discoveries has been significantly bolstered by the deployment of both terrestrial observatories and orbital missions, notably the Kepler Space Telescope, whose vast contributions have decisively shaped our current methodologies for detecting exoplanets (Borucki et al., 2010).

Building upon these observational milestones, our research seeks to probe the extensive datasets available, with a focused examination of patterns and relationships among exoplanets located within their host stars' habitable zones (HZs). The concept of planetary habitability is central to astrophysics and astrobiology, particularly concerning the identification of planets that may support life. Various studies have articulated criteria for habitability, primarily emphasizing the necessity of surface-located liquid water, suitable atmospheric conditions, and orbital stability (Kasting et al., 1993; Seager et al., 2007). Expanding on this conceptual framework, our study delves into critical parameters such as surface liquid water potential, orbital dynamics, and stellar luminosity, each recognized as pivotal in assessing a planet's potential to harbor life (Kasting, 1988; Lammer et al., 2009).

The definition of the habitable zone has undergone substantial evolution over several decades, transitioning from a simplistic model based on radiative flux to a more comprehensive framework that integrates various stellar and planetary characteristics. The foundational models by Huang (1959) and Hart (1979) introduced the habitable zone as the orbital region where a planet could maintain liquid water on its surface. This definition has been refined over time to account for complex variables such as orbital eccentricity and stellar luminosity fluctuations, culminating in concepts such as the Continuously Habitable Zone (CHZ) (Kasting & Catling, 2003; Kopparapu et al., 2013).

Recent research has also explored additional determinants of habitability, such as planetary atmospheres, axial tilt, and magnetic fields, which further complicate the habitability assessment (Lammer et al., 2009; Shields et al., 2016). While recognizing the significance of these broader parameters, our analysis remains focused on the more conventionally measurable factors due to the current limitations of exoplanetary data and the desire to maintain analytical clarity and consistency.

Moreover, our study acknowledges the dynamic nature of habitable zones, which can shift due to the processes of stellar evolution. Such changes pose significant implications for long-term planetary habitability, as demonstrated in studies investigating the impact of stellar luminosity variations over time (Kasting & Catling, 2003; Rushby et al., 2013).

This paper is structured to first provide a comprehensive review of the prevailing theories and empirical findings concerning habitable zones and the established criteria for planetary habitability. Following this, we describe our methodology, detailing the criteria used to delineate potential habitable zone exoplanets and the standardized metrics for our calculations. We then present our findings regarding the distribution and specific characteristics of these exoplanets, thereby contributing novel insights into their potential habitability. Finally, we discuss the broader implications of our results for the field of exoplanetary science and propose future research directions that may further illuminate the complex interplay of factors influencing exoplanetary habitability.

## 2. Methodology

The primary dataset was sourced from the NASA Exoplanet Archive, a comprehensive repository of exoplanet data. The focus was on the 5,595 confirmed exoplanets listed in the Archive's "Planetary Systems Composite Data" section as of March 10th, 2024. This dataset was organized to facilitate the efficient application of computational analyses and the interpretation of complex patterns. Observational specifics such as exoplanet orbital parameters and host star radiative characteristics were systematically accounted for. The approach leverages detailed descriptions within the dataset, enabling the extraction of statistically significant relationships based chiefly on primary observables such as host star mass, type, and exoplanet semi-major axis.

2.1 Addressing Observational Biases

Acknowledging and accounting for the biases inherent in common exoplanet detection methods is crucial to enabling the extraction of meaningful results and conclusions. We critically analyzed how the Transit method, predominantly detecting short-period exoplanets, and the Radial Velocity method, favoring massive exoplanets, might skew our dataset. To counter these biases and refine our dataset's reliability, we considered using exclusively Kepler mission data, known for its comprehensive and diverse exoplanet discoveries. While this strategic selection aimed to provide a more balanced view of the exoplanet population, excluding non-Kepler derived data would introduce its own set of biases simply through the absence of valuable confirmed exoplanetary data from other sourcing. Alternatively, quantitative contrasts were drawn between the population distribution of stellar classes in general throughout the Milky Way galaxy and those comprising host stars contained in the Archive. Further, our analysis focused on single (known) host exoplanetary systems in order to provide consistency in considering the key habitability factors such as radiative environments and orbital mechanics.

2.2 Habitability Zone Determination

The habitable zone of a given exoplanet was determined based on its calculated average surface temperature, denoted as $T_{surf,ave}$, a critical indicator of potential liquid water presence. We categorized exoplanets as "Too Hot" ($T_{surf,ave}$ >100°C), "Too Cold" ($T_{surf,ave}$ < 0°C) or within the habitable zone ("In HZ") for $T_{surf,ave}$ between the benchmark temperature range of 0 to 100ºC. This classification was vital to identifying exoplanets that could potentially support life under the assumed necessary precursor of surface-accessible liquid phase $H_2O$.

Utilizing the basic Radiative Equilibrium equation as derived from first principles of radiative heat transfer, the exoplanet average surface temperature calculation considered several characterizing factors. These include the exoplanet's distance from its host star ($d$), the host's effective surface temperature ($T_\odot$) and radius ($R_\odot$), exoplanet albedo (A) and an additional scaler to account for bulk atmospheric greenhouse gas effect ($k$):

$$T_{surf,ave} = kT_\odot(1-A)^{0.25}(R_\odot/(2d))^{0.5} \qquad (1)$$

Applying equation (1) to each listing in the NASA Exoplanet Archive enabled selective sifting of the database to produce quantified results based on the aforementioned exoplanet HZ status categories. Additionally, where the Archive had no entry for the observable parameters $T_\odot$ and/or $R_\odot$ and/or $d$, a designation of "N/A" was made for the associated exoplanet to denote insufficient information for determining HZ status in those cases.

2.3 Stellar Classification and Exoplanet Distribution

We examined how exoplanetary formation correlates with various stellar classes. This analysis involved sorting habitable zone exoplanets by their host stars' spectral types and investigating the variations in HZ boundary distances in relation to host star mass. This part of our methodology was designed to uncover trends in exoplanet distribution across different types of stars, providing insights into the likelihood of habitable planets around various stellar classes. Further elucidation of aforementioned observational bias was sought by challenging conventional assumptions regarding the propensity for hosting planetary systems based on stellar type.

2.4 Non-Applicable and Additional Factors

The study acknowledged but did not focus on several complexing factors which can influence habitability, such as orbital eccentricity, axial obliquity, atmospheric composition, and the presence (or lack thereof) of exoplanetary magnetic fields and/or exomoons, due to data limitations. Some insights into the temporal variability of habitable zones – i.e., the impact of stellar evolution on the habitability of orbiting planets over time – can be deduced qualitatively from statistical treatment of the data.

2.5 Computational Approach

The "Planetary Systems Composite Data" from the NASA Exoplanet Archive as of March 10, 2024, was the primary dataset, chosen for its consolidation of multiple observations for a given exoplanet into single line data entries. Custom calculations for thermal equilibrium and habitable zone estimations were added directly into the resultant spreadsheet. This enhanced the functionality of the spreadsheet by folding-in analytical tool capability, enabling extraction of meaningful patterns and relationships from the nearly 5,600 confirmed exoplanet entries. Additional relevant data, such as host star age, was also incorporated to further extend the analysis.

2.6 Assumptions and Limitations

Our analysis was underpinned by several assumptions, notably adopting Earth's albedo (A = 0.306) as a baseline for exoplanets as well as accounting for the atmospheric greenhouse gas effect through the bulk temperature factor (k = 1.13), again using Earth as the standard. Recognizing that our empirical relationships, based on a large but nonetheless limited dataset, might introduce certain biases, we were careful to frame our findings within these constraints. Where assumptions were necessary to complete calculations, rational bracketing conditions were applied accordingly.

## 3. Results

Analysis of exoplanet habitability within circumstellar habitable zones reveals several critical insights, visualized through a series of figures that illustrate various aspects of the data. By linking these figures together, a narrative is constructed that enhances the understanding of exoplanetary habitability and aligns it with existing literature. Systematic examination of 5,595 confirmed exoplanets from the NASA Exoplanet Archive, applying equation (1) to calculate the average surface temperatures ($T_{surf,ave}$) and categorize their habitable zone status, was performed.

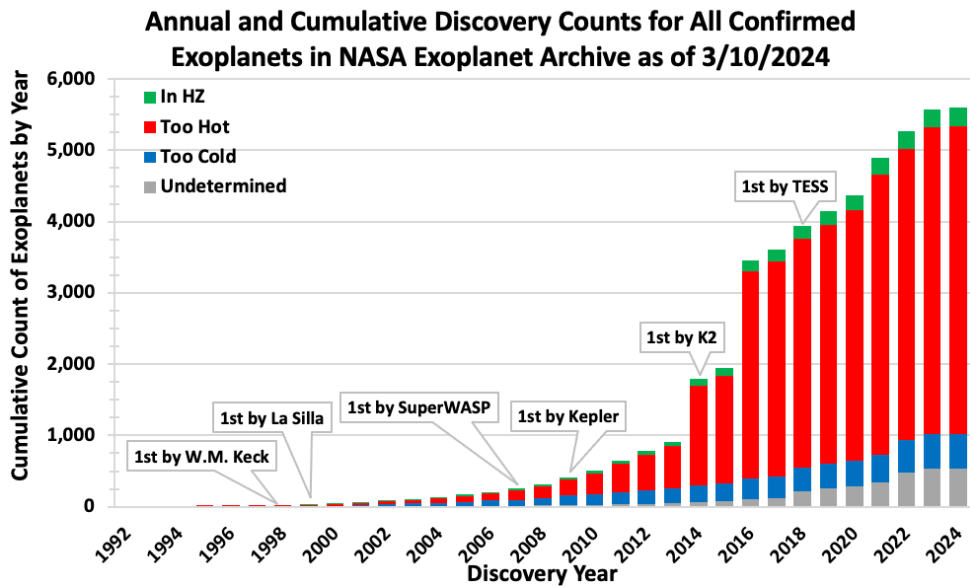

**Figure 1:** Cumulative count history of all confirmed exoplanet discoveries showing habitable zone status per the chosen parameters of this analysis. Note: Data for 2024 is incomplete.

Figure 1 depicts the cumulative count history of all confirmed exoplanet discoveries, highlighting their habitable zone status. The exponential growth in exoplanet discoveries, particularly since the launch of the Kepler Space Telescope, underscores significant advancements in detection technologies and methodologies (Borucki et al., 2010). This trend illustrates our increasing ability to identify potentially habitable exoplanets, reflecting ongoing refinements in search strategies and expanding observational capabilities. The incomplete data for 2024 is indicative of ongoing discoveries, suggesting that the counts of habitable and non-habitable zone exoplanets continued growth, driven by new technologies enabling more advanced data analysis techniques and observation missions.

To understand the spatial distribution of exoplanet discoveries, Figure 2 illustrates the distance-wise distribution of all confirmed exoplanets from the Solar System, categorized by their habitable zone status. The majority of exoplanets are concentrated within 1,000 light-years, highlighting observational biases where nearer stars are more frequently surveyed due to limitations in current detection technologies. This bias underscores the challenge of detecting distant exoplanets and emphasizes the need for next-generation telescopes capable of probing deeper into the galaxy to identify more distant, potentially habitable exoplanets (Gaudi et al., 2020). The clustering of discoveries within 1,000 light-years also reflects the limitations in the sensitivity and resolution of current instruments, as well as the prioritization of closer stars for detailed observation (Winn & Fabrycky, 2015).

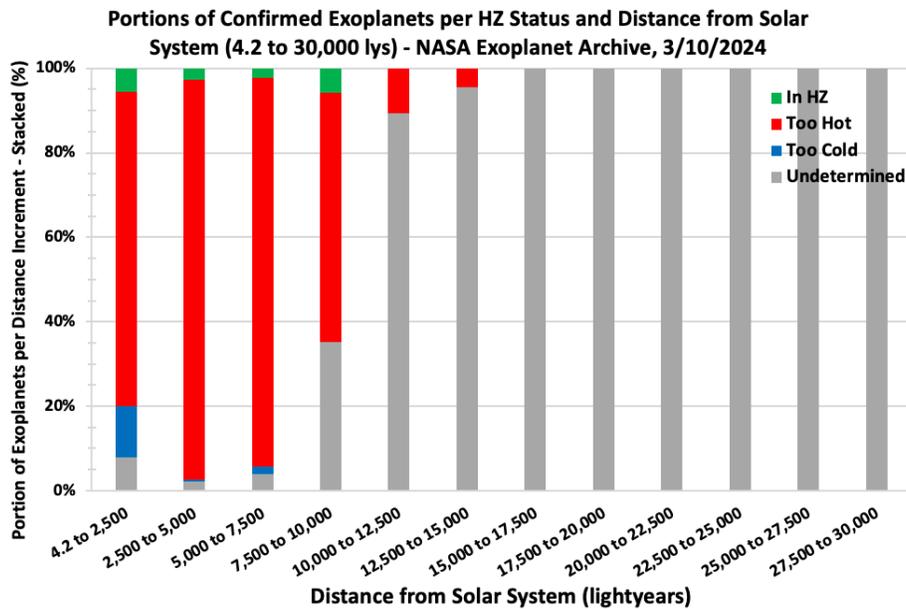

**Figure 2:** Distance-wise distribution of all confirmed exoplanets from Solar System and their habitable zone status per the chosen parameters of this analysis.

Figure 3 presents the distribution of confirmed single-hosted exoplanets based on their habitable zone status. Among these exoplanets, 77.75% are categorized as "Too Hot," 8.04% as "Too Cold," 4.48% are within the HZ, and 9.73% have indeterminate status (N/A). The significant proportion of "Too Hot" exoplanets suggests an observational bias, as closer-in planets with shorter orbital periods are easier to detect using methods like the Transit method. This finding aligns with previous studies that highlight the detection bias towards short-period exoplanets, often leading to an underrepresentation of planets within the habitable zone (Seager & Mallén-Ornelas, 2003). The skew towards "Too Hot" exoplanets also reflects the challenges in detecting cooler, potentially habitable planets that lie farther from their host stars, where longer orbital periods and lower transit probabilities complicate their detection (Howard et al., 2012).

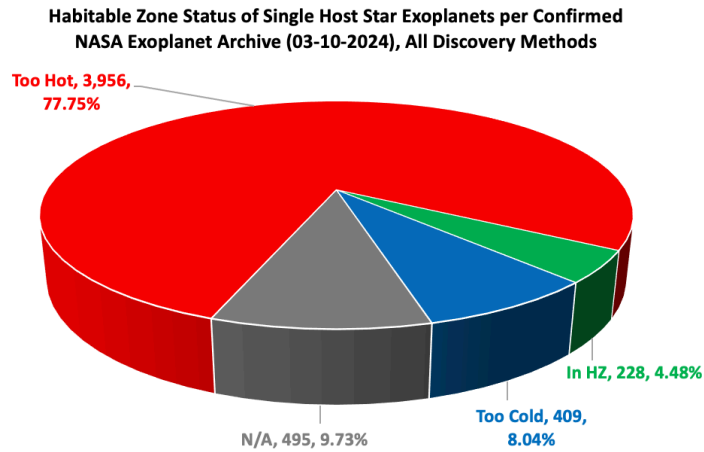

**Figure 3:** Count-up and portion of confirmed single-hosted exoplanets' habitable zone status per the chosen parameters of this analysis.

Delving deeper into the detection methods, Figure 4 (a) shows the habitable zone status of single-hosted exoplanets discovered via the Transit and Transit Timing Variations methods. Specifically, 89.75% are "Too Hot," 0.90% are "Too Cold," 3.10% are within the HZ, and 6.25% are N/A. The overwhelming majority of "Too Hot" exoplanets discovered through these methods underscores the inherent observational bias towards detecting planets with shorter orbital periods, which are more likely to transit their host star frequently. This bias reflects the limitations of the Transit method in identifying habitable zone exoplanets, as it is more sensitive to planets that orbit closer to their stars (Petigura et al., 2013). The underrepresentation of "Too Cold" and HZ planets emphasizes the need for more sensitive instruments and extended observation periods to detect planets in wider orbits.

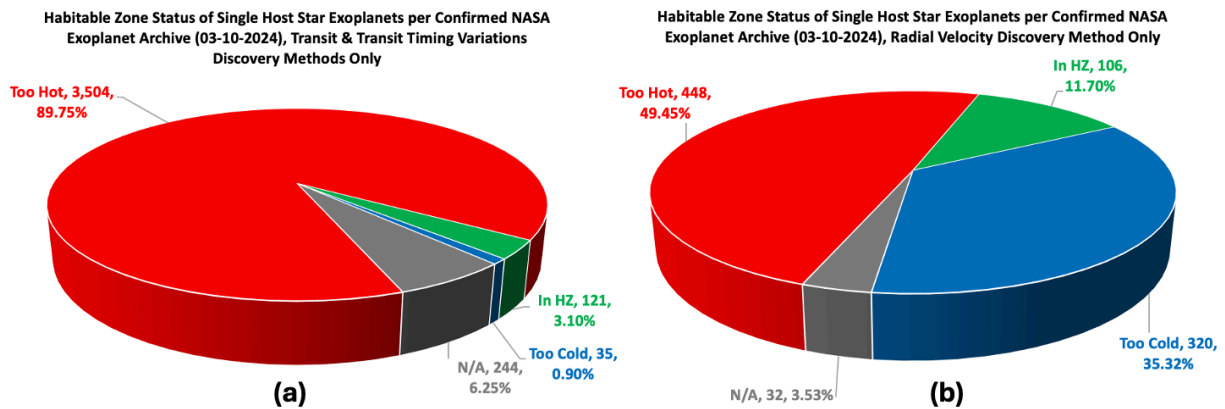

**Figure 4:** (a) Left: Habitable zone status of single hosted exoplanets discovered by way of the Transit and Transit Timing Variations methods; (b) Right: Habitable zone status of single hosted exoplanets discovered by way of the Radial Velocity method.

Figure 4 (b) presents the habitable zone status of single-hosted exoplanets discovered using the Radial Velocity method. The distribution is more balanced compared to the Transit method, with 49.45% "Too Hot," 35.32% "Too Cold," 11.70% within the HZ, and 3.53% N/A. The Radial Velocity method's sensitivity to planets at various distances from their host stars provides a broader view of exoplanetary systems, although it still shows a detection bias towards larger planets. This method's ability to detect planets in a wider range of orbits highlights its complementary role in identifying habitable zone exoplanets, addressing some of the limitations inherent in the Transit method (Mayor & Queloz, 1995). The broader distribution of HZ exoplanets detected by Radial Velocity indicates its potential in revealing more distant, possibly habitable planets that are missed by transit surveys. Note: Appendix-Figure 1 extends the analysis presented in Figure 4 (a) and 4 (b) to the Imaging method, revealing this technique's detection capability for HZ exoplanets to still be quite limited in comparison.

Moving to the stellar classifications of host stars, Figure 5 shows the proportions of host stars across all single-hosted exo-systems containing at least one HZ exoplanet. The breakdown is as follows: 26.32% are M-type stars, 29.82% are K-type stars, 35.53% are G-type stars, and 7.89% are F-type stars. This distribution indicates a higher prevalence of HZ exoplanets around G-type and K-type stars, aligning with the fact that these stars are prime targets for habitability studies due to their stable lifetimes and favorable conditions for liquid water (Kasting et al., 1993). The relatively lower proportion of HZ exoplanets around M-type stars, despite their abundance in the galaxy, reflects the challenges in detecting potentially habitable planets around these dimmer, cooler stars (Shields et al., 2016). M-type stars, while abundant, have smaller HZs closer to the star, making planets within these zones more susceptible to stellar activity and tidal locking, which could hinder habitability (Dressing & Charbonneau, 2015).

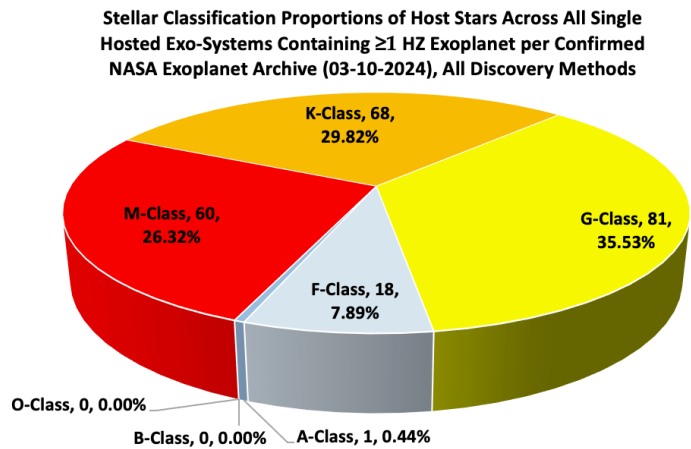

**Figure 5:** Host stellar class proportions for all single host star exo-systems which contain at least one habitable zone exoplanet. Note: Single hosts of more than one HZ exoplanet are counted multiply according to HZ exoplanet count in those exo-systems.

Figure 6 compares the relative abundance of M, K, G, F, A, and B stellar classes in the Milky Way galaxy to the stellar classes of confirmed exoplanet host stars in single-host systems. For exoplanet host stars, the proportions are 7.8% M-type, 24.7% K-type, 47.4% G-type, 19.4% F-type, 0.6% A-type, and 0.2% B-type. In contrast, the overall abundance in the Milky Way is 76.5% M-type, 12.1% K-type, 7.6% G-type, 3.0% F-type, 0.6% A-type, and 0.13% B-type. This discrepancy highlights a selection bias towards G-type stars, which are similar to our Sun and are often prioritized in exoplanet searches due to their potential for habitability (Reid et al., 2002). The underrepresentation of M-type star hosts further underscores the observational challenges and biases in current exoplanet surveys. The preference for G-type stars also reflects historical biases and the assumption that solar analogs are more likely to host habitable planets (Brown, 2015).

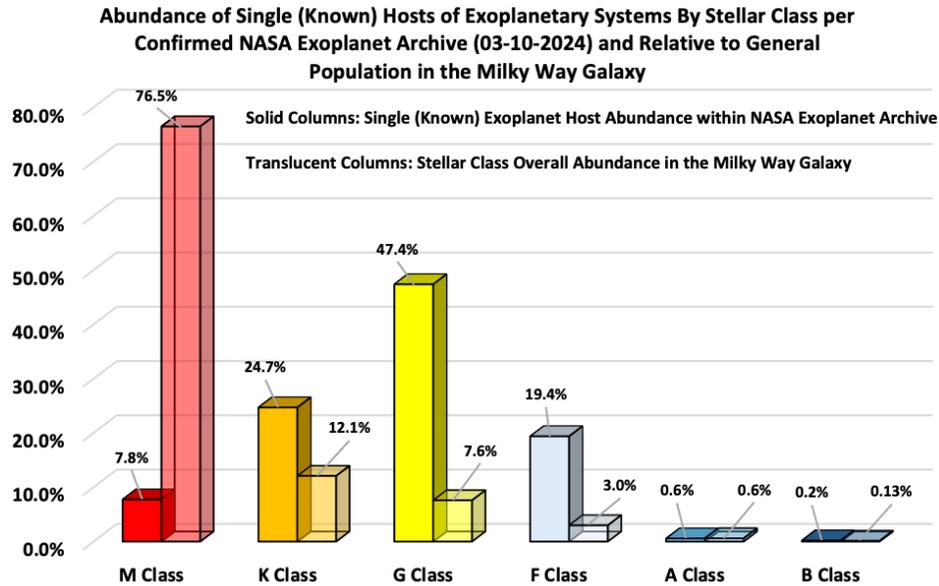

**Figure 6:** Relative abundance of M, K, G, F, A & B stellar classes in the Milky Way galaxy to the stellar classes of confirmed exoplanet host stars of single host systems. Note: The host star of each singly-hosted exoplanetary system is counted only once, regardless of how many exoplanets in that system are hosted or their HZ status.

Examining exoplanet discovery methods further, Figure 7 shows the proportion of single-hosted exoplanets categorized by these various methodologies. The breakdown is as follows: Transit 76.20%, Radial Velocity 17.81%, Pulsation Time Variation 0.04%, Pulsar Timing 0.12%, Orbital Brightness Modulation 0.18%, Microlensing 3.95%, Imaging 1.12%, Astrometry 0.04%, Disk Kinematics 0.02%, and Transit Timing Variations 0.53%.

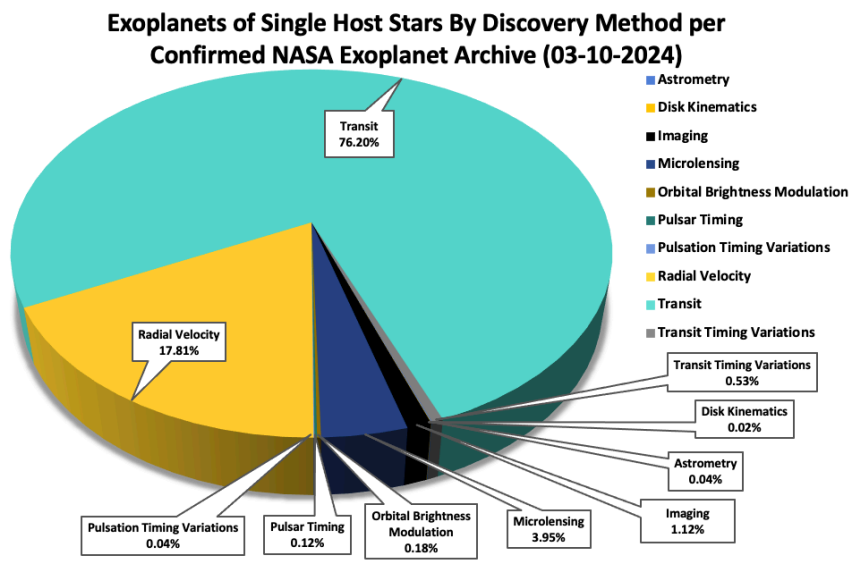

**Figure 7:** Single-hosted exoplanets portioned by discovery method.

The dominance of the Transit method reflects its efficiency in detecting exoplanets, particularly those close to their host stars. This method's predominance in exoplanet discovery highlights its strengths in surveying large areas of the sky and detecting numerous exoplanets, although it also emphasizes the need for complementary methods to provide a more complete picture of exoplanetary systems (Winn & Fabrycky, 2015). The significant presence of exoplanets discovered via microlensing and imaging methods showcases their role in identifying planets at greater distances and in different stellar environments (Bennett et al., 2014). Note: Appendix-Figure 2 focuses the analysis of Figure 7 specifically to HZ exoplanets, illustrating the co-dominance and near parity in terms of overall counts between the Transit and Radial Velocity methods.

Figure 8 shows the habitable zone width as a function of host star mass, including only those hosts with mass known to ≤10% uncertainty. The Sun is included for reference as a large yellow dot. The figure indicates that the width of the habitable zone increases with the mass of the host star and can be approximated as a power function. This relationship aligns with theoretical models where more massive stars have broader habitable zones due to their higher luminosities, which affect the range of distances at which liquid water could exist on a planet's surface (Kasting et al., 1993). This finding suggests that more massive stars may offer wider potential zones for habitability, though they also present challenges such as shorter lifespans and higher levels of stellar activity (Selsis et al., 2007). The broader HZ around massive stars implies that planets can orbit at greater distances while still maintaining surface conditions conducive to liquid water, but the increased stellar radiation and shorter stellar lifetimes may limit long-term habitability (Kopparapu et al., 2013).

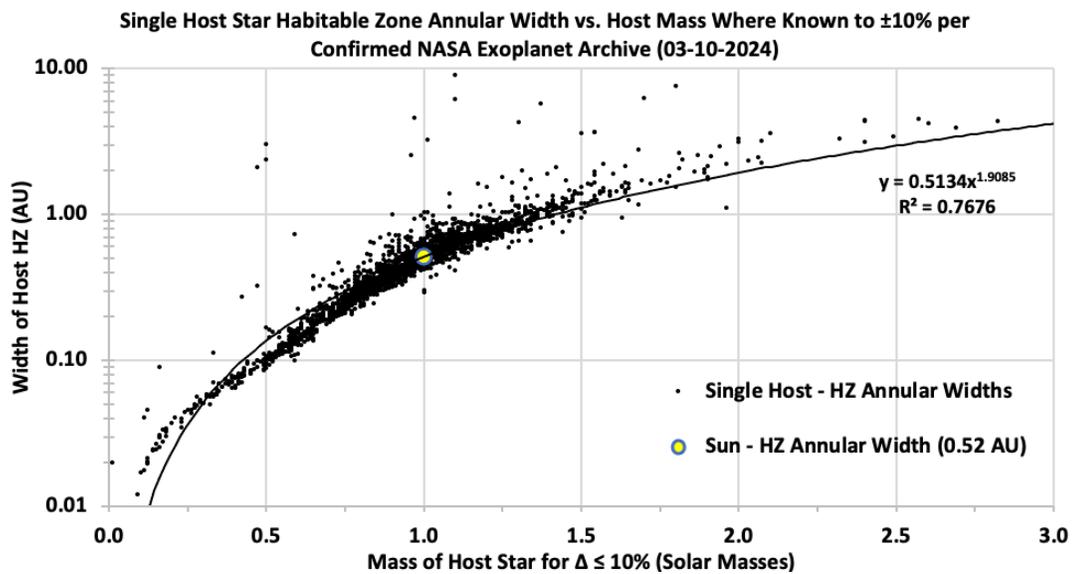

**Figure 8:** Habitable zone width of exoplanet single host stars as a function of host mass, includes only hosts with mass known to ≤ 10% uncertainty. Sun (large yellow dot) included for reference.

Figure 9 illustrates the variation of habitable zone boundary distances from the host star as a function of host mass, overlaid with exoplanet semi-major axes corresponding to host mass. The inner (red dots) and outer (blue dots) boundaries of the HZ are shown alongside the semi-major axes of exoplanets. The Solar System's planets, as well as the Sun's similarly calculated HZ inner and outer HZ boundaries, are depicted for reference. This visualization demonstrates how the HZ boundaries expand outward with increasing host star mass, while the distribution of exoplanet semi-major axes suggests a tendency for planets to reside closer to their stars in lower-mass systems and further away in higher-mass systems. This pattern reflects the influence of stellar mass on planetary formation and orbital dynamics, providing insights into the distribution of potentially habitable exoplanets across different stellar environments (Kopparapu et al., 2013). The data suggest that in lower-mass systems, planets are more likely to be found closer to the star, within the narrower HZ, while in higher-mass systems, planets can be situated further out within the broader HZ, potentially offering a more stable environment for life (Lammer et al., 2009).

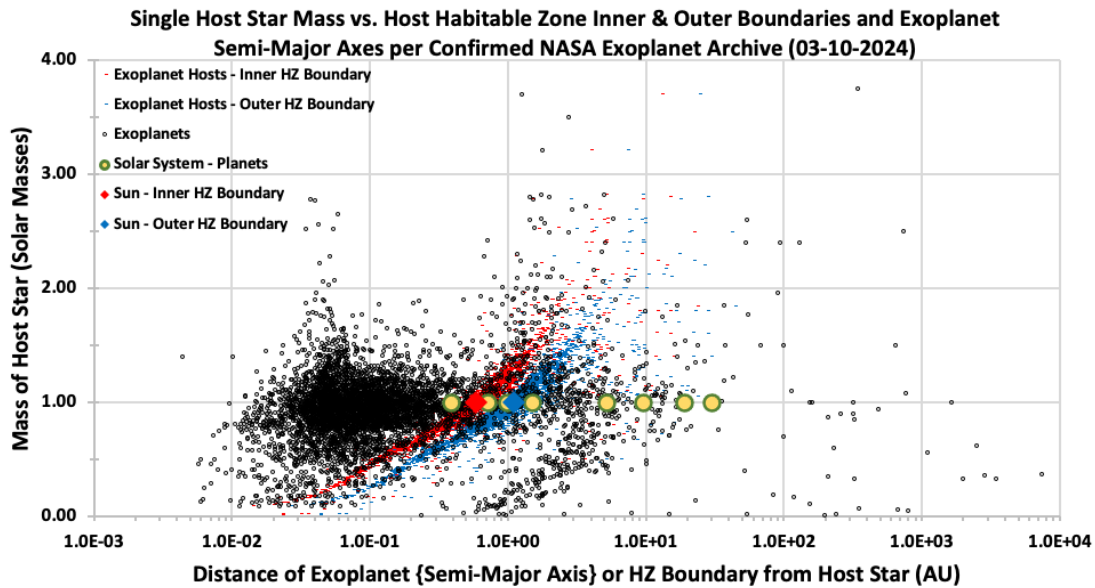

**Figure 9:** Variation of habitable zones' boundary distances from host star for single hosts as functions of host mass, overlaid with exoplanet semi-major axis and corresponding host mass.

Figure 10 illustrates the relationship between the effective temperatures of single host stars and the surface temperatures of their corresponding exoplanets, with exoplanets grouped by type: Gas Giants, Neptunian Planets, Super-Earths, and Terrestrial Planets. The Earth's position is labeled for reference, along with those for the Solar System's other planets, with the habitable zone range indicated within the vertical green dashed lines. The size of each point represents the relative size of the exoplanet. An exoplanetary system of particular note is TRAPPIST-1, consisting of seven confirmed planets and the host itself, an ultra-cool (effective surface temperature of 2,566 K) red dwarf much smaller in radius than the Sun. Notwithstanding, the TRAPPIST-1 system contains two HZ planets and three of terrestrial size, one of which is among the HZ pair as calculated using equation (1). Given its intriguing array of exoplanets, this system has been one of the more closely studied since its discovery in 2016.

This figure provides a comprehensive visualization of the relationship between stellar temperatures and exoplanet surface temperatures. It shows that few exoplanets fall within the HZ range, indicating the rarity of conditions suitable for surface liquid water. Higher stellar temperatures generally correspond to higher planetary surface temperatures, as evident from the upward trend of data points. Gas Giants and Neptunian Planets predominantly lie outside the HZ, while Super-Earths and Terrestrial Planets show a greater propensity to occupy or approach the HZ. This visualization underscores the need for advanced observational technologies and methodologies to discover and characterize exoplanets within the HZ. Future missions equipped with direct imaging capabilities and improved sensitivity are essential to identifying and studying Earth-like planets in habitable zones of a broader range of stellar types.

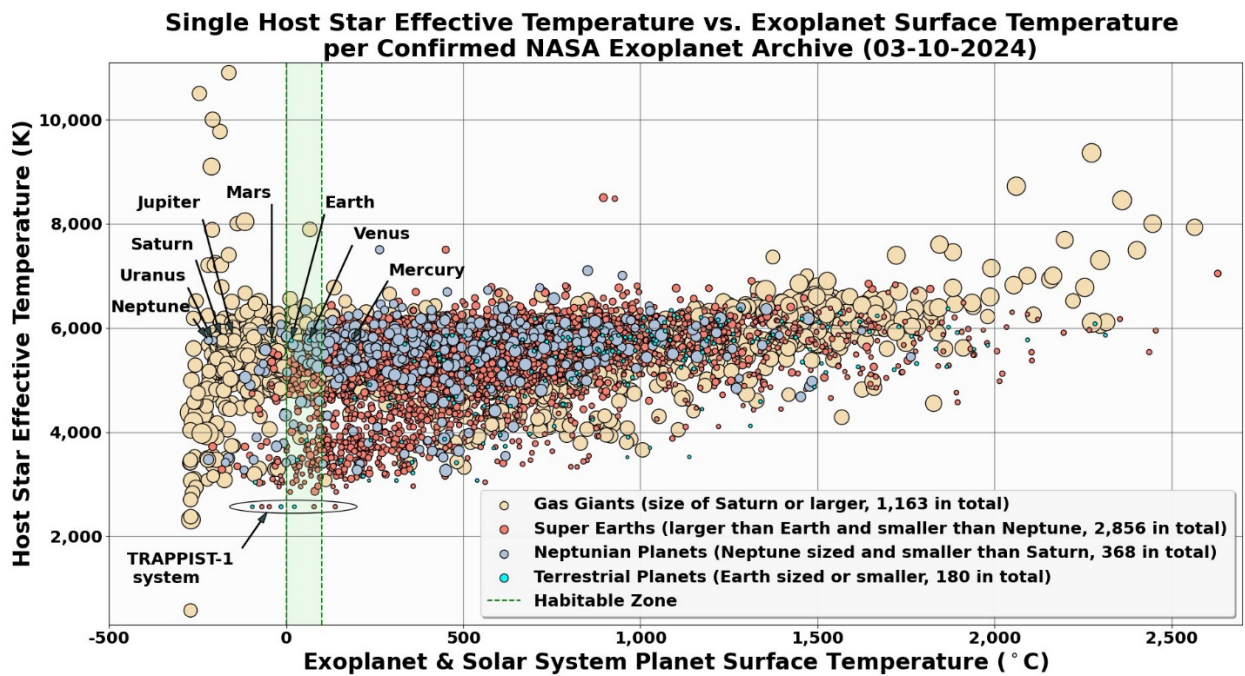

**Figure 10:** Single Host Star Effective Temperature vs. Exoplanet & Earth Surface Temperature. This scatter plot shows the effective temperature of single host stars versus the surface temperature of their corresponding exoplanets. Exoplanets are grouped by types: Gas Giants (size of Saturn or larger), Neptunian Planets (Neptune-sized, smaller than Saturn), Super-Earths (larger than Earth, smaller than Neptune), and Terrestrial Planets (Earth-sized or smaller). The positions of the Earth and all other solar system planets are labeled for reference, along with the habitable zone (HZ) range indicated by green dashed lines. The size of each point represents the relative size of the exoplanet. Note: The seven known planets of the TRAPPIST-1 system are indicated at lower left.

This figure also brings attention to assumption limitations introduced earlier in this paper. For the planets in our solar system, surface temperatures calculated with Equation (1) generally align with known mean temperatures (https://science.nasa.gov/resource/solar-system-temperatures/) with a 6-37% difference. However, Venus is an exception. Although its surface is too hot for life as we know it, Equation (1) flags the planet as within the habitable zone. This discrepancy arises from the assumption of a standardized bulk temperature factor (k=1.13k) based on Earth's values when accounting for the atmospheric greenhouse effect. In reality, Venus has a very thick atmosphere composed primarily of $CO_2$, trapping heat and resulting in a much higher bulk temperature factor (k=3.17). This limitation is discussed earlier in section 2.6. Bracketing the inner Solar System-based atmospheric greenhouse assumption, this on the cooler end, is Mars. While the atmosphere of Mars is also predominantly composed of $CO_2$, it is far less dense and accordingly much less capable of trapping solar radiation. The particular exception of Venus indicates that variations in the atmospheric greenhouse effect will need to be further considered to better determine exoplanet surface temperatures.

Overall, the analysis presented above highlights several observational biases inherent in current exoplanet detection methods. The Transit method, which dominates exoplanet discoveries, is more likely to detect planets with shorter orbital periods, leading to an overrepresentation of "Too Hot" exoplanets. This bias is evident in the high percentage of "Too Hot" planets discovered via Transit methods (Figures 3 and 4a). Conversely, the Radial Velocity method, which can detect exoplanets at various distances from their host stars, presents a more balanced distribution of habitable zone statuses (Figure 4b).

The stellar classification of host stars (Figures 5 and 6) shows a preference for G-type stars in exoplanet searches, despite M-type stars being by far the most common in the Milky Way. This selection bias may be due to the more stable and longer-lived nature of G-type stars, which are conducive to sustaining life-supporting environments over extended periods.

Figures 8 and 9 provide insights into the relationship between host star mass and habitable zone characteristics. The widening of the habitable zone with increasing host star mass suggests that more massive stars offer a larger range of distances where conditions might support surface liquid water. However, the semi-major axis distribution of exoplanets indicates a concentration of planets closer to lower-mass stars, potentially due to the higher likelihood of detecting such planets through current observational techniques – again echoing the dominance of the Transit method.

Figure 10 adds additional context by illustrating the relationship between the effective temperatures of host stars and the surface temperatures of their corresponding exoplanets. This figure highlights the challenges in finding planets within the HZ and underscores the importance of advanced observational technologies to overcome current limitations.

Overall, these results emphasize the need for continued development and deployment of diverse detection methods to achieve a more comprehensive understanding of exoplanetary systems. Future missions should aim to mitigate observational biases by targeting a broader range of stellar types and distances, thereby enhancing our ability to identify potentially habitable exoplanets and build a more complete understanding of planetary systems in general.

## 4. Summary and Discussion

The investigation into the habitability of exoplanets within circumstellar habitable zones (HZs) has provided valuable insights and contributed to the broader understanding of exoplanetary science. By analyzing 5,595 confirmed exoplanets from the NASA Exoplanet Archive, this study has highlighted several key findings that advance our knowledge of potential habitable worlds beyond our Solar System.

The exponential growth in exoplanet discoveries underscores the advancements in detection technologies and methodologies, especially with contributions from missions such as the Kepler/K2 Space Telescope and the Transiting Exoplanet Survey Satellite (TESS). This growth trend indicates that our capability to identify exoplanets, with an ongoing emphasis on those orbiting within their host HZ, will continue to improve with future technological advancements in both data gathering missions and analysis.

One of the significant findings from this study is the importance of addressing observational biases inherent in different exoplanet detection methods. The Transit method, while efficient in detecting numerous exoplanets, shows a bias towards closer-in, hotter planets. In contrast, the Radial Velocity method provides a more balanced view but still favors larger planets. This highlights the need for employing a combination of detection methods to achieve a comprehensive understanding of exoplanet populations and to identify potentially habitable exoplanets more effectively.

The study also emphasizes the crucial role of stellar classifications in determining exoplanet habitability. G-type and K-type stars are shown to host a higher proportion of habitable zone exoplanets, aligning with their stable lifetimes and favorable conditions for surface liquid water. However, the underrepresentation of M-type stars, despite their general abundance, points to the challenges and potential habitability issues associated with these stars, such as stellar activity and tidal locking.

Moreover, the analysis of habitable zone widths as a function of host star mass reveals that more massive stars have broader habitable zones due to their higher luminosities. This suggests that such stars may offer wider potential zones for habitability, though their shorter lifespans and higher levels of stellar activity could pose challenges for long-term habitability. This finding aligns with theoretical models and provides a deeper understanding of how stellar characteristics influence planetary habitability.

The discrepancy between the stellar class distribution of exoplanet host stars and that of the Milky Way's general distribution indicates a selection bias towards Sun-like G-type stars. This historical focus on solar analogs for habitability studies underscores the necessity of broadening survey efforts to include a wider range of stellar types, thus providing a more comprehensive view of potential habitable planets across different host star environments.

Despite robust theoretical modeling, more thorough analysis of exoplanetary systems—and particularly the potential habitability of planets therein—requires additional detailed observational data. As Ramirez (2018) suggests, "this situation would vastly improve with technological (e.g., engineering) advancements in observational techniques, including direct imaging. Even with current limitations, we have suggested some observations that can still be made with upcoming and next generation missions." It is vital to continue studying data that encompasses the broad spectrum of possibilities.

Another aspiration of more in-depth analysis is the unification of the Circumstellar Habitable Zone, Galactic Habitable Zone, and Cosmic Habitable Age concepts. The ultimate goal is to formulate a comprehensive methodology for determining if any strongly Earth-like planets exist, such as the one proposed by Cai et al. (2021), which suggests that planets most likely to harbor life are located in an annular region approximately 4 kpc from the galactic center. Integrating each piece of the puzzle leads us towards a better understanding of where life might exist in the galaxy. The years ahead are expected to prove particularly enlightening as more powerful ground and space-based instrumentation comes online, along with advancements in computational techniques for determining the habitability of worlds outside the Solar System.

In conclusion, this study enhances existing models for managing burgeoning exoplanetary datasets and lays foundational groundwork for future explorations into the dynamic relationships between exoplanets and their stellar environments. By addressing observational biases, employing diverse detection methods, and considering the influence of stellar classifications, this research contributes to a more nuanced understanding of exoplanet habitability. The findings underscore the importance of multifaceted approaches and enhanced observational capabilities in the ongoing search for extraterrestrial life. Future research will benefit from these insights, guiding efforts to uncover the full spectrum of potentially habitable exoplanets and advancing our quest to understand our place in the cosmos.

## Acknowledgement:

This research was conducted at the NASA sponsored Jet Propulsion Laboratory, California Institute of Technology (Caltech). It has made use of the NASA Exoplanet Archive, which is operated by Caltech, under contract with the National Aeronautics and Space Administration under the Exoplanet Exploration Program. The authors also thank the supports from the Pacific Academy at Irvine, California, and Lakeside School at Seattle, Washington.

## Data Statement:

The data underlying this article can be downloaded from the NASA exoplanet archive at https://exoplanetarchive.ipac.caltech.edu. The method of data calculation and analysis are fully described in the article.

## References:

Bennett, D. P., et al. (2014). "The Discovery of an Exoplanet in the Microlensing Event OGLE-2012-BLG-0563." The Astrophysical Journal, 785(2), 155.

Borucki, W. J., et al. (2010). "Kepler Planet-Detection Mission: Introduction and First Results." Science, 327(5968), 977-980.

Brown, A. G. A. (2015). "The Gaia Mission: Science, Organization and Present Status." Proceedings of the International Astronomical Union, 11(A29A), 29-38.

Cai, M. X., et al. (2021). "Searching for Life in the Galactic Habitable Zone." The Astrophysical Journal, 921(2), 87.

Dressing, C. D., & Charbonneau, D. (2015). "The Occurrence of Potentially Habitable Planets Orbiting M Dwarfs Estimated from the Full Kepler Dataset and an Empirical Measurement of the Detection Sensitivity." The Astrophysical Journal, 807(1), 45.

Gaudi, B. S., et al. (2020). "The Habitable Exoplanet Observatory (HabEx) Mission Concept Study Report." arXiv preprint arXiv:2001.06683.

Gonzalez, G. (2005). "Habitable Zones in the Universe." Origins of Life and Evolution of the Biosphere, 35(6), 555-606.

Hart, M. H. (1979). "Habitable Zones about Main Sequence Stars." Icarus, 37(1), 351-357.

Howard, A. W., et al. (2012). "Planet Occurrence within 0.25 AU of Solar-type Stars from Kepler." The Astrophysical Journal Supplement Series, 201(2), 15.

Huang, S. S. (1959). "Occurrence of Life in the Universe." American Scientist, 47(3), 397-402.

Kasting, J. F. (1988). "Runaway and Moist Greenhouse Atmospheres and the Evolution of Earth and Venus." Icarus, 74(3), 472-494.

Kasting, J. F., & Catling, D. (2003). "Evolution of a Habitable Planet." Annual Review of Astronomy and Astrophysics, 41, 429-463.

Kasting, J. F., Whitmire, D. P., & Reynolds, R. T. (1993). "Habitable Zones around Main Sequence Stars." Icarus, 101(1), 108-128.

Kopparapu, R. K., et al. (2013). "Habitable Zones around Main-sequence Stars: New Estimates." The Astrophysical Journal, 765(2), 131.

Lammer, H., et al. (2009). "What Makes a Planet Habitable?" The Astronomy and Astrophysics Review, 17(2), 181-249.

Mayor, M., & Queloz, D. (1995). "A Jupiter-mass Companion to a Solar-type Star." Nature, 378(6555), 355-359.

NASA Exoplanet Archive (2023). "Catalog of Confirmed Exoplanets."

Petigura, E. A., et al. (2013). "Prevalence of Earth-size Planets Orbiting Sun-like Stars." Proceedings of the National Academy of Sciences, 110(48), 19273-19278.

Ramirez, R. M. (2018). "Exploring Exoplanet Surface Habitability at Different Stellar Energies." Nature Astronomy, 2(1), 32-37.

Reid, I. N., et al. (2002). "Meeting the Cool Neighbors. VIII. A Preliminary 20 PC Census from the NStars Database." The Astronomical Journal, 124(5), 2721-2738.

Rushby, A. J., et al. (2013). "Habitable Zone Lifetimes of Exoplanets around Main Sequence Stars." Astrobiology, 13(9), 833-849.

Seager, S., & Mallén-Ornelas, G. (2003). "A Unique Solution of Planet and Star Parameters from an Extrasolar Planet Transit Light Curve." The Astrophysical Journal, 585(2), 1038-1055.

Seager, S., et al. (2007). "Mass-Radius Relationships for Solid Exoplanets." The Astrophysical Journal, 669(2), 1279-1297.

Selsis, F., et al. (2007). "Habitable Planets around the Star Gliese 581?" Astronomy & Astrophysics, 476(3), 1373-1387.

Shields, A. L., et al. (2016). "The Effect of Orbital Configuration on the Possible Climates and Habitability of Kepler-62f." Astrobiology, 16(6), 443-464.

Shields, A. L., et al. (2016). "The Habitability of Planets Orbiting M-dwarf Stars." Physics Reports, 663, 1-38.

Winn, J. N., & Fabrycky, D. C. (2015). "The Occurrence and Architecture of Exoplanetary Systems." Annual Review of Astronomy and Astrophysics, 53, 409-447.